\pdfoutput=1 

\documentclass[journal]{IEEEtran}
\usepackage{multirow} 

\usepackage{cite}

\ifCLASSINFOpdf
   \usepackage[pdftex]{graphicx} 
\else
\fi
\usepackage{amsmath}
\ifCLASSOPTIONcompsoc
  \usepackage[caption=false,font=normalsize,labelfont=sf,textfont=sf]{subfig}
\else
  \usepackage[caption=false,font=footnotesize]{subfig}
\fi
\begin{document}
%
\title{Capacity-achieving and Flicker-free FEC Coding Scheme for Dimmable Visible Light Communication Based on Polar Codes}
%
%
%

\author{Junbin~Fang, Zhen Che, Xiaolong Yu, Zhe Chen, Zoe L. Jiang$\dagger$, Siu-Ming~Yiu, Kui Ren,  Xiaoqing Tan
\thanks{J. Fang and Z. Che are joint first authors, they contribute equally to the paper.}
\thanks{J. Fang, X. Yu and Z. Chen are with the Department of Optoelectronic Engineering, and the Key Laboratory of Visible Light Communications of Guangzhou, Jinan University, Guangzhou, 510632, China. J. Fang is also with The Edward S. Rogers Sr. Department of Electrical \& Computer Engineering Department, University of Toronto.}
\thanks{Z. Che is with the College of Information Science and Technology, Jinan University, Guangzhou, 510632, China.}
\thanks{$\dagger$Zoe L. Jiang is with Harbin Institute of Technology Shenzhen Graduate School, Shenzhen 518055, China. e-mail: zoeljiang@gmail.com}
\thanks{S.M. Yiu is with the Department of Computer Science, The University of Hong Kong, HKSAR, China.}
\thanks{K. Ren is with the Department of Computer Science and Engineering, State University of New York at Buffalo (SUNY Buffalo), USA.}
\thanks{X. Tan is with the Department of Mathematics, Jinan University, Guangzhou, 510632, China.}
\thanks{This work has been submitted to the IEEE for possible publication.  Copyright may be transferred without notice, after which this version may no longer be accessible.}
}

\maketitle

\begin{abstract}
Visible light communication (VLC) could provide short-range optical wireless communication together with illumination using LED lighting. 
However, conventional forward error correction (FEC) codes for reliable communication do not have the features for dimming support and flicker mitigation which are required in VLC for the main functionality of lighting. 
Therefore, auxiliary coding techniques are usually needed, which eventually reduce the coding efficiency and increase the complexity. 
In this paper, a polar codes-based FEC coding scheme for dimmable VLC is proposed to increase the coding efficiency and simplify the coding structure.  
Experimental results show that the proposed scheme has the following advantages: 1) equal probability of 1's and 0's in codewords, which is inherently supporting $50\%$ dimming balance; 2) short run length property (about $90\%$ bits have runs shorter than 5) which can avoid flickers and additional run-length limited line coding; 3) higher coding efficiency about twofold than that of other coding schemes; 4) capacity achieving error correction performance with low-complexity encoding and decoding, which is about 3 dB higher coding gain than that of RS(64,32) in IEEE standard for dimming ratio $50\%$ and about 1 dB higher coding gain than that of LDPC codes for dimming ratio $25\%$ (or $75\%$).
\end{abstract}


%
\IEEEpeerreviewmaketitle

%
%
%
%
\section{Introduction}
\label{sec_intro}

\IEEEPARstart{V}{isible} light communication (VLC) has attracted increasing attention as a novel short-range wireless communication technology since it can provide both illumination and communication service simultaneously~\cite{jovicic2013visible,o2008visible,komine2004fundamental}, especially with the widespread deployment of LED lighting for energy saving.
VLC is being considered as a supplementary method for future fifth-generation (5G) wireless communications~\cite{chow2015enhancement,takai2014optical} also due to its wide bandwidth and immunization to electromagnetic interference.
Since VLC uses on-off keying intensity modulation (OOK IM) to convey digital information into LED light signal, the brightness and the stabilization of light are affected by the distribution of 1's and 0's in data frames as the LEDs are turned on or off dependent on the data bits being 1 or 0.
Therefore, forward error correction (FEC) coding scheme for VLC has two main challenges, dimming support and flicker mitigation, to maintain the functionality and the stability of illumination~\cite{rajagopal2012ieee,lee2015modulation}. 

FEC codes usually do not guarantee the equal probability of 1's and 0's (ON and OFF symbols), causing light intensity unbalance and requiring additional adjustments to support arbitrary dimming ratio.
Besides, FEC codes which cannot avoid long runs of 1's and 0's may cause light intensity fluctuation over a short period, which may exceed the persistence of the human eye and causes noticeable brightness change, i.e., flicker.
In IEEE VLC standard~\cite{ieee2011std}, dimming function is supported via inserting compensation symbols (CSs) for OOK modulation to achieve desired average intensity of LED light.
To mitigate the flickers, run-length limited (RLL) line coding, such as Manchester codes and etc., is suggested to break long runs of 1's and 0's~\cite{rajagopal2012ieee}.
However, the inserted CSs and RLL codes significantly reduce the overall coding efficiency and the bit transmission rate.
For example, Manchester code is a rate-1/2 code which encodes one bit into two symbols. 
Besides, these auxiliary coding procedures complicate the coding structure of VLC system and introduce extra latency.

Several modified FEC coding schemes considering flicker mitigation and dimming support have been proposed to improve the overall coding efficiency and the error correction performance~\cite{kim2011novel,kim2013modified,lee2012turbo,kim2015adaptive,kim2014acoding,lu2016achieving,feng2015fountain}.
However, most of these coding schemes cannot guarantee DC balance and short runs of 1's and 0's.
Additional coding techniques are still required, such as code puncturing, scrambling, RLL line codes and etc.
As a consequence, even advanced coding schemes such as low-density parity check (LDPC) and turbo codes are used~\cite{lee2012turbo, kim2015adaptive}, the coding overhead is still relatively high, resulting in low coding efficiency with complicated coding structure.
Besides, none of these codes are proven to achieve Shannon capacity for a binary symmetric channel (BSC) and binary input additive white Gaussian noise channel (BIAWGNC).

In this paper, a capacity-achieving and flicker-free FEC coding scheme using polar codes is proposed for dimmable VLC.
Polar codes, recently introduced by Arikan~\cite{arikan2009channel}, are a family of codes that provably achieve the capacity of symmetric binary memoryless channels with ``low encoding and decoding complexity" in the order of $O(NlogN)$, where $N$ is the length of the codewords.
Taking advantages of polar codes' recursive channel combination structure, the proposed FEC coding scheme could not only achieve a higher coding efficiency than the other schemes but also provide dimming support and flicker mitigation functionalities without auxiliary coding procedures, which significantly simplifies the coding structure for dimmable VLC system.
Experimental results show that the proposed scheme has the following advantages: 1) equal probability of 1's and 0's in codewords, which is inherently supporting $50\%$ dimming balance; 2) short run length property which can avoid flickers and additional run-length limited line coding; 3) higher coding efficiency about two-fold than that of the other coding schemes; capacity achieving error correction performance with low-complexity encoding and decoding, which is about 3 dB higher coding gain than that of RS(64,32) in IEEE standard for dimming ratio $50\%$ and about 1 dB higher coding gain than that of LDPC codes for dimming ratio $25\%$ (or $75\%$).

The rest of this paper is organized as follows.
Section~\ref{sec_related} reviews and compares previous FEC coding schemes for dimmable VLC.
The details of the proposed FEC coding scheme are introduced in Section~\ref{sec_scheme}. 
The experimental results are shown and discussed in Section~\ref{sec_results}.
Section~\ref{sec_conclusion} concludes the paper.

\section{Related Works}
\label{sec_related}

\begin{figure}[!t]
\centering
\includegraphics[width=2.5in]{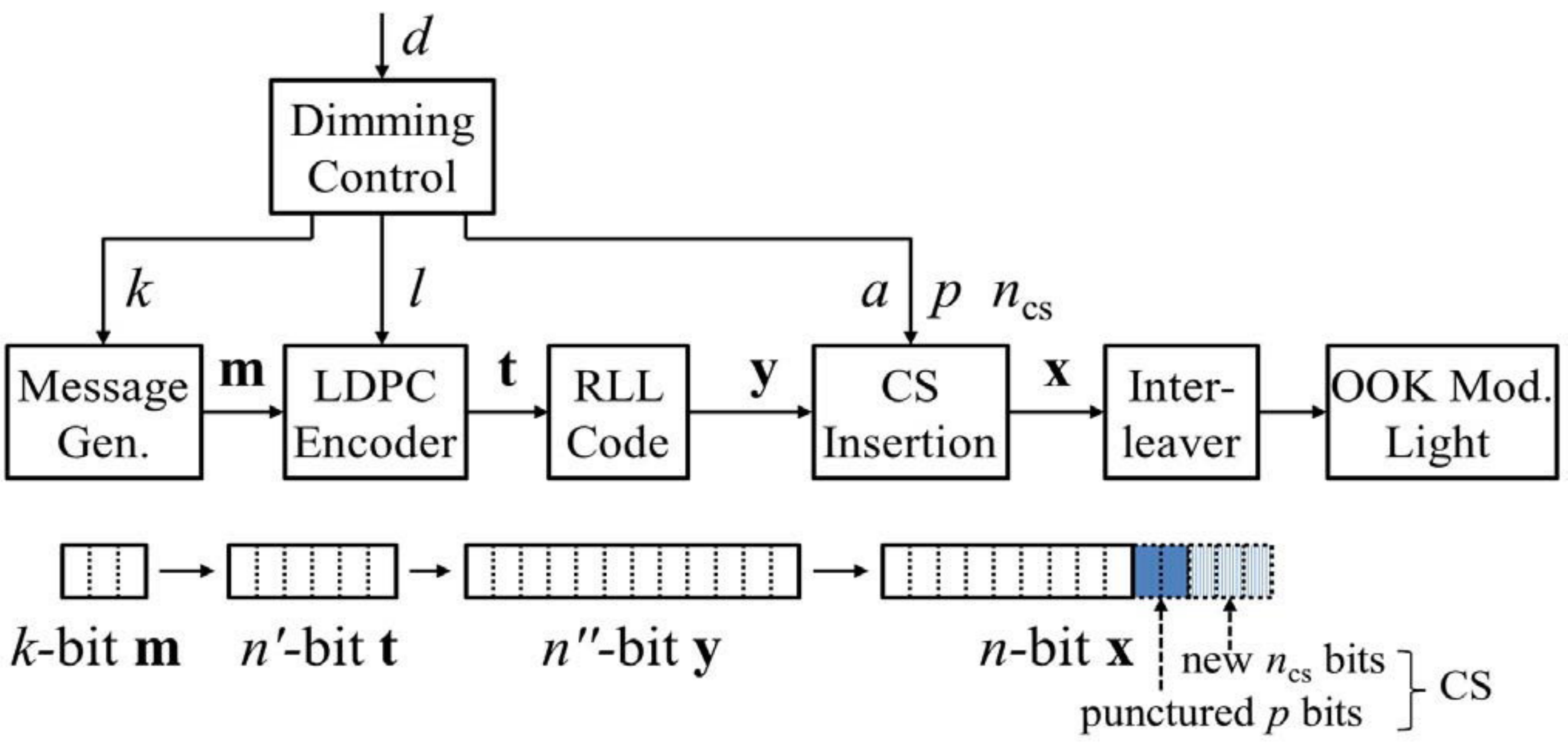}
\caption{Typical coding structure for dimmable VLC.}
\label{fig_oldscheme}
\end{figure}

From the reliable communication point of view, FEC codes are indispensable for VLC.
However, as the convergence of illumination and communication, VLC not only aims at data transmission but also concerns two lighting related problems: flicker and dimming~\cite{rajagopal2012ieee}.
To avoid flicker, the changes in brightness must fall within the maximum flickering time period (MFTP), defined as the maximum time period over which the light intensity can change without the human eye perceiving it. 
In general, a frequency greater than 200 Hz (MFTP $<$ 5 ms) is generally considered safe~\cite{berman1991human}. 
Dimming requirement is a practical issue as users may adjust the LED light intensity according to the needs of the current environment and themselves.
For OOK modulation, VLC dimming can be achieved by inserting dummy symbols and the average duty cycle of the waveform can be changed by the insertion of ``compensation" time into the modulation waveform within the data frame while sacrificing data rate.
Therefore, a typical coding structure for dimmable VLC is illustrated in Figure~\ref{fig_oldscheme}, which is borrowed from Ref.~\cite{kim2015adaptive}.
In this example, message $m$ is first encoded by LDPC encoder, and then the codeword is further encoded via RLL line coding to break long runs of 1's or 0's to erase possible flickers.
Finally, some CSs are inserted to adjust the ratio of ``ON" time to ``OFF" time to achieve the targeted dimming ratio.

In IEEE VLC standard~\cite{ieee2011std, rajagopal2012ieee}, either a Reed-Solomon (RS) code or a concatenated code composed of RS and convolutional codes (CC) is directly used as the error-correcting code (ECC) for short data frames.
No modification of FEC codes is made in the scheme as it uses CSs for dimming function and RLL line codes for flicker mitigation. 
In recent years, several improved coding schemes have been proposed~\cite{kim2011novel,kim2013modified,lee2012turbo,kim2015adaptive,kim2014acoding,lu2016achieving,feng2015fountain}. 

Two coding schemes based on modified Reed-Muller (RM) codes and minimal use of CSs have been proposed by Kim and Jung for dimming support in VLC systems~\cite{kim2011novel,kim2013modified}. 
Although the RM codes-based schemes can guarantee the DC balance at exact 50\%, they cannot achieve high code rate as their code rates scale with $O( {log_2(N)/N} )$ for codeword length $N$, and their error correction performances are also inferior to iterative decoding codes such as turbo or LDPC codes.

To overcome the limitation of RM codes-based scheme, two kinds of coding schemes using advanced error correction codes were proposed.
The first one is a turbo codes-based coding scheme proposed by Lee and Kwon~\cite{lee2012turbo}. 
However, it requires code puncturing, pseudo-noise sequence scrambling and CSs inserting to avoid long runs of 1s or 0s and to match the codewords' weight with the desired dimming ratio, introducing high coding overhead.
Besides, for the dimming ratios greater than $1/2$, all the binary symbols of the codewords need to be flipped to
obtain those dimming ratios.
Otherwise, the puncturing rate will be too high and make the information in a message indistinguishable.

The other one is an LDPC codes-based scheme proposed by Kim to adaptively adjust dimming values in VLC systems~\cite{kim2015adaptive}. 
As shown in Figure~\ref{fig_oldscheme}, its dimming function is also provided via puncturing methods and CSs, and RLL codes are used to avoid long runs.
The significant advantages of the method include the lower number of codes required to support codes with different coding rate and the relatively small performance degradation for dimming control.
Nevertheless, with the rate-1/2 LDPC codes, the effect of the puncturing rate and the rate-1/2  Manchester line codes, the overall code rate could be very low. 
For example, a 3-bit message will be encoded into a 15-bit codeword using the LDPC-based scheme.
Furthermore, these two schemes require iterative decoding algorithm, which increases the computational cost and latency.

The similar problems also exist in the joint FEC-RLL coding mechanism proposed by Lu and Li trying to simultaneously achieve bandwidth efficiency, error correction and run-length control~\cite{lu2016achieving}.
The concatenated convolutional-Miller coding scheme consists of an OOK modulated Miller code serving as the inner code and a convolutional code serving as the outer code. 
Unfortunately, Miller codes have disappointing power efficiency and bit error rate (BER) performance, which cannot be ignored in practice.


A fountain codes-based encoding scheme was also proposed by Feng et. al.~\cite{feng2015fountain} to improve transmission efficiency and used the least CSs.
However, the coding scheme requires a feedback link, which may not be available in VLC broadcasting scenario and introduces constraint to the whole system.
Beside, scrambling is also needed.

In a brief summary, the existing FEC coding schemes for dimmable VLC system have similar problems such as low coding efficiency and complicated coding structure. 
Table~\ref{tab_overview} gives a brief overview and comparison of these schemes including the proposed scheme.

\begin{table}[!t]
\renewcommand{\arraystretch}{1.3}
\caption{An overview on FEC coding schemes for dimmable VLC}
\label{tab_overview}
\centering
\begin{tabular}{|c|c|c|}
\hline
Schemes & Dimming support& Flicker mitigation \\
\hline
RS codes(IEEE)~\cite{ieee2011std} & CS Insertion& RLL line codes\\
\hline
RM codes~\cite{kim2011novel,kim2013modified}  & CS Insertion& Not mentioned\\
\hline
Turbo codes~\cite{lee2012turbo} & Puncturing & Scrambling\\
\hline
LDPC codes~\cite{kim2015adaptive}  & CS Insertion + Puncturing& RLL line codes \\
\hline
Fountain codes~\cite{feng2015fountain}  &CS Insertion& Scrambling\\ 
\hline
CC-RLL~\cite{lu2016achieving} & CS Insertion& Miller codes\\
\hline
The proposed scheme & CS Insertion & No need\\
\hline
\end{tabular}
\end{table}

\section{Dimmable VLC System with Polar Codes}
\label{sec_scheme}
\subsection{VLC System Model}
\label{sec_model}

\begin{figure}[!t]
\centering
\includegraphics[width=2.5in]{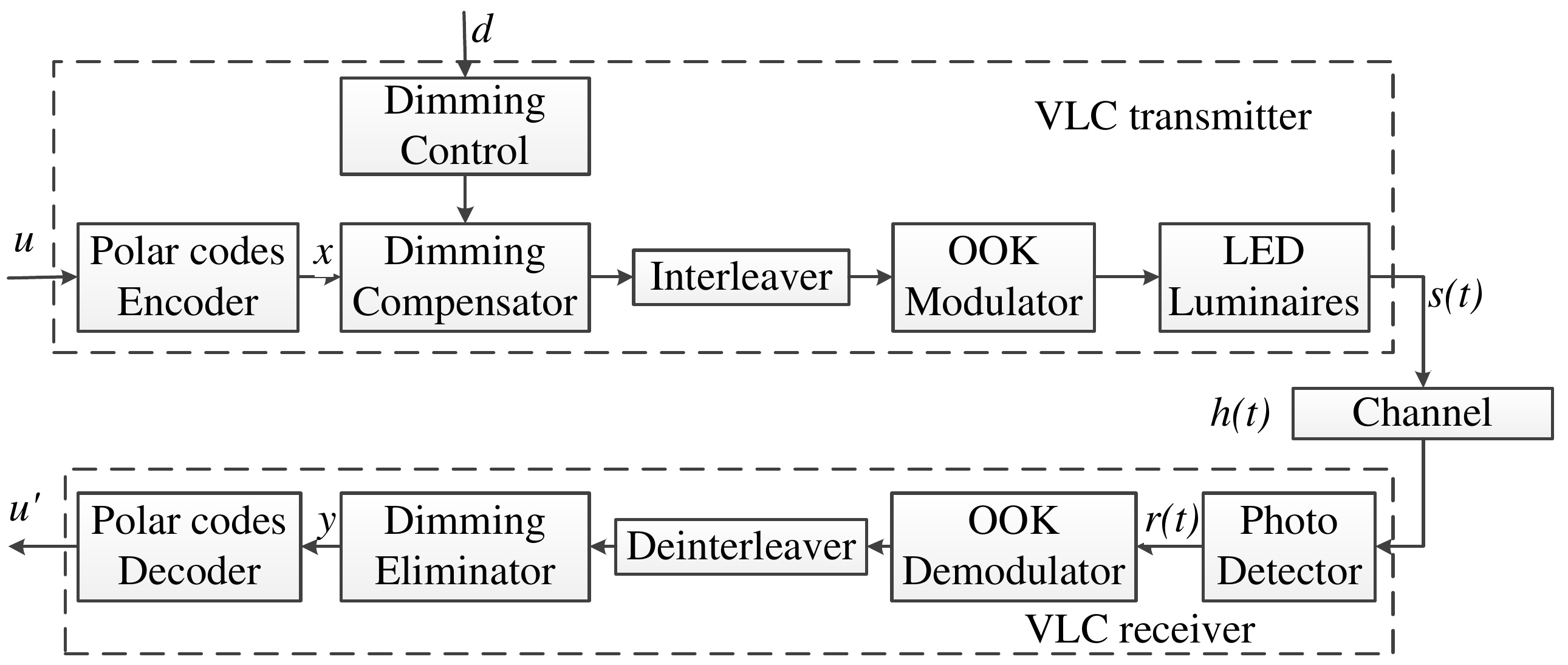} 
\caption{Block diagram of the proposed VLC system.}
\label{fig_system}
\end{figure}

Block diagram of dimmable VLC system with the proposed coding structure is shown in Figure~\ref{fig_system}. 
At the transmitting side, information vector $\mathbf{u}$ with $k$ bits is input into the polar codes encoder to generate encoded codeword $\mathbf{x}$ with length $N$, and then the following dimming compensator inserts appropriate number of CSs into the data frame according to dimming control to achieve desired dimming ratio. 
The dimming ratio, $d$, corresponds to the ratio of the number of ``on"  OOK symbols over data frame's length, e.g., $d = 50\%$ means that the numbers of on-bit and off-bit are equal to half of the data frame's length.
Denote the number of the inserted dimming CSs as $N_{cs}$. Then, the length of the transmitted data frame is $N_{frame}=N+N_{cs}$.
Therefore, the code rate of the FEC codes is $R_c=k/N$ and the overall coding efficiency is given by $\eta_{overall}=k/N_{frame}$.
Next, an interleaver is utilized for permuting codeword bits and compensation symbol bits to mitigate the effect of channel fading which may make burst errors in received data frames.
After the interleaving and OOK modulating processes, the interleaved codeword will be emitted via LED luminaires as light signal $s(t)$, which has the average optical power
$P_t(=(1/T)\int_{0}^{T}s(t)dt)$, where $T$ denotes the light signal duration.

After passing through the VLC channel $h(t)$, the visible light signal is received by a photodiode (PD). Then the received signal is given as $r(t)=R \cdot x(t) \ast h(t)+n(t)$ \cite{komine2009adaptive, kahn1997wireless}, where ``$\ast$" denotes the convolution, $R$ is the PD conversion efficiency, and $n(t)$ is the additive white Gaussian noise (AWGN) which contains the shot, thermal noise and intersymbol interference. 
After the OOK demodulating and deinterleaving processes, the polar codes decoder estimates the transmitted information and outputs the decoded message $\mathbf{\hat{u}}$. 

\subsection{FEC coding Scheme based on Polar Codes}
\label{sec_proposed}
Polar codes were introduced by Erdal Arikan in 2009 and they provide the first deterministic construction of capacity-achieving codes for symmetric binary-input discrete memoryless
channels (B-DMCs)~\cite{arikan2009channel}.
Their code construction is based on a phenomenon called ``channel polarization", with which individual copies of B-DMCs are combined recursively in order to form a new set of channels composed of more and more differentiated channels, such that in the asymptotic limit channels are either error-free or completely noisy, with a proportion of error-free channels equal to the channel capacity.
A detailed description and proof of polar codes can be found in~\cite{arikan2009channel}.

\subsubsection{Channel Polarization}
\label{sec_construct}
In particular, consider a generic B-DMC $W:\mathcal{X} \to \mathcal{Y}$ with input alphabet $\mathcal{X}$, output alphabet $\mathcal{Y}$ and transition probabilities $W(y|x), x\in\mathcal{X}, y \in\mathcal{Y}$. The symmetric capacity of this channel $W$ is denoted as $I(W)$, i.e., the highest rate at which reliable communication is possible. 
In general, transmitting $N$ bits via channel $W$ could be regarded as $N$ independent uses of $W$, denoted by $W^N:\mathcal{X}^N \to \mathcal{Y}^N$ with $W^N(y_1^N|_1^N)=\prod_{i=1}^NW(y_i|x_i)$.

Polar codes use $W^N$ in a different way, channel polarization. Channel polarization consists of a channel combining phase and a channel splitting phase.
First, $N$ independent copies of $W$ are combined in a recursive manner to produce a \emph{vector channel} $W_N:\mathcal{X}^N \to \mathcal{Y}^N$.
In fact, this process is to build up correlations among the $N$ channels through $n(=logN)$ stages of $N/2$ interleaved binary combinations: 
$W_2(y_1, y_2|u_1,u_2)=W(y_1|x_1)W(y_2|x_2)=W(y_1|u_1\oplus u_2)W(y_2|u_2)$.
After that, $W_N$ are split into $N$ binary-input \emph{coordinate channels} $W_N^{(i)}:\mathcal{X} \to \mathcal{Y}^N\times\mathcal{X}^{i-1}, 1 \leq i \leq N$ that show a polarization effect in the sense that, as $N$ becomes large, the symmetric capacity terms ${I(W_N^{(i)})}$ tend towards 0 or 1 for all but a vanishing fraction of indices $i$.

Therefore, information bits could be designed to be transmitted via the noiseless coordinate bit-channels for which ${I(W_N^{(i)})}$ is near 1, and frozen bits with preset values could be designed to be transmitted via the pure noisy coordinate bit-channels for which ${I(W_N^{(i)})}$ is near 0, so as to improve the transmission rate and the reliability of the system. 
Denote $N_{info}$ as the number of capacity near 1 channels and $N_{frozen}$ as that of capacity near 0 channels, then we have $N=N_{info}+N_{frozen}$, and the code rate could be up to $R_{polar}=N_{info}/N=I(W)$.
Note that in this paper, $N$ equals to $N_c$, the length of the encoded codewords $x$ excluding CSs.


\subsubsection{Polar Encoder}
\label{sec_encoder}
For a given block length $N$, a message $u$ is encoded into codeword $x$ in the manner, namely $x_1^N=u_1^NG_N$, where $u_1^N$ is the information bit with index from $1$ to $N$ and $G_N$ is the generator matrix of order $N$, which is given as $G_N=B_NF^{\otimes n}$, where 
$F= \left[ \begin{array}{cc}
    1 & 0 \\
    1 & 1 
  \end{array}
\right]$ and $n=log_2(N)$;
$``\otimes``$ denotes Kronecker product; $B_N$ denotes the $N\times N$  bit reversal permutation matrix. 

For $\mathcal{A}$ an arbitrary subset of $\{1,\ldots,\emph{N}\}$, the codeword can be written as 
$ 
x_1^N=u_\mathcal{A}G_N(\mathcal{A}) \oplus u_{\mathcal{A}^c}G_N(\mathcal{A}^c)
$
where $G_N(\mathcal{A})$ denotes the submatrix of $G_N$ formed by the rows with indices in $\mathcal{A}$; $\mathcal{A}^c$ is the complement of the subset $\mathcal{A}$ in set $\{1,\ldots,\emph{N}\}$; $``\oplus"$ denotes XOR operation. 
$\mathcal{A}$ and $\mathcal{A}^c$ are referred as the information set and the frozen set, i.e., the optimized coordinate channel set and the degraded coordinate channel set, respectively.
Correspondingly, $u_{\mathcal{A}}$ and $u_{\mathcal{A}^c}$ refer to information bits and frozen bits.
In general, the complexity of this encoding algorithm is $O(NlogN)$ with $O(N)$ for $B_N$ and $O(NlogN)$ for $F^{\otimes n}$.

\subsubsection{Polar Decoder}
\label{sec_decoder}
After the encoded codeword $x_1^N$ is sent over the channel $W^N$, a channel output $y_1^N$ will be received. 
The decoder's task is to generate an estimate $\hat{u}_1^N$ of $u_1^N$, given knowledge of $\mathcal{A}$, $u_{\mathcal{A}^c}$ (here are all 0's) and $y_1^N$. 
Since the bit values of frozen part were preseted, the decoder can simply set $\hat{u}_{\mathcal{A}^c}=u_{\mathcal{A}^c}$ to eliminate the errors in this part and use these known bits to help the real decoding task, generating an estimated $\hat{u}_{\mathcal{A}}$ of $u_{\mathcal{A}}$ .

Polar codes decoding has a recursive structure similar to that of encoding.
A recursive successive-cancellation (SC) decoder observes $(y_1^N,u_{A^c})$ and computes the likelihood ratio (LR) of $\hat{u}_1^N$ bit by bit from index $1$ to $N$ using the recursive formulas given as
\begin{displaymath}
L_N^{(2i-1)}(y_1^N,\hat{u}_1^{2i-2})
=\frac{
L_{N/2}^{(i)}(y_1^{N/2},\hat{u}_{1,o}^{2i-2} \oplus \hat{u}_{1,e}^{2i-2})L_{N/2}^{(i)}(y_{N/2+1}^N,\hat{u}_{1,e}^{2i-2})+1}
{ 
L_{N/2}^{(i)}(y_1^{N/2},\hat{u}_{1,o}^{2i-2} \oplus \hat{u}_{1,e}^{2i-2})+L_{N/2}^{(i)}(y_{N/2+1}^N,\hat{u}_{1,e}^{2i-2})
}
\end{displaymath}
and
\begin{displaymath}
L_N^{(2i)}(y_1^N,\hat{u}_1^{2i-1})
=\left[ 
L_{N/2}^{(i)}
\left( y_1^{N/2},\hat{u}_{1,o}^{2i-2} \oplus \hat{u}_{1,e}^{2i-2}
\right)
\right] ^{1-2 \hat{u}_{2i-1}}
\cdot
L_{N/2}^{(i)}(y_{N/2+1}^N,\hat{u}_{1.e}^{2i-2})
\end{displaymath}
and decides the bit value by hard decision,
where $\hat{u}_1^{2i-2}$ and $\hat{u}_1^{2i-1}$ mean the previously estimated bits which have indices before the bit under computation, $\hat{u}_{1,o}^{2i-2}$ and $\hat{u}_{1,e}^{2i-2}$ mean the estimated bits with odd indices and even indices, respectively.

The above formulas show that computing LR of bits in $N$-length polar codes could be decomposed into recursively calling the polar decoding algorithm for two $N/2$-length polar transforms defined by $I_2\otimes G_{N/2}$, until the length $N/2$ reaches 1.
Besides, the LR computation for $\hat{u}^i$ will not be activated until the decoder finishes the previous decisions $\hat{u}_1^{i-1}$.
The complexity of this decoding algorithm is determined essentially by the complexity of computing the LRs, which is $O(NlogN)$ \cite{arikan2009channel}.

\section{Experimental Results and Discussion}
\label{sec_results}
A series of experiments were conducted to exemplify the performance of the proposed scheme in terms of code weight distribution, run length property, coding efficiency and error correction.
In concrete, the code weight distribution test illustrates the even probabilities of the existences of 0's and 1's guaranteed by the proposed scheme, even at different (arbitrary) code rates, and therefore reflects the dimming support inherently provided by the coding scheme without additional coding techniques or constraints.
Secondly, the run length test indicates that the coding scheme holds short run length features which can mitigate lighting flicker without the help of RLL line coding.
Based on the above two advantages, the proposed coding scheme could be simplified and achieves a higher overall coding efficiency than the other existing schemes.
Finally, the error correction test shows that the coding scheme has outstanding BER performance and better coding gains, compared with modified RM codes~\cite{kim2011novel}, RS(64,32) codes in IEEE standard 802.15.7~\cite{ieee2011std}, and LDPC codes~\cite{kim2015adaptive}.

Without loss of generality, the length of the codewords is 1024 bits in these experiments, i.e., 1024 symbols in OOK modulation.

\subsection{Code Weight Balance for Dimming Support}
\label{sec_codeweight}
\begin{figure}[!t]
\centering
\subfloat[Code rate $R_c=1/4$]{\includegraphics[width=2.5in]{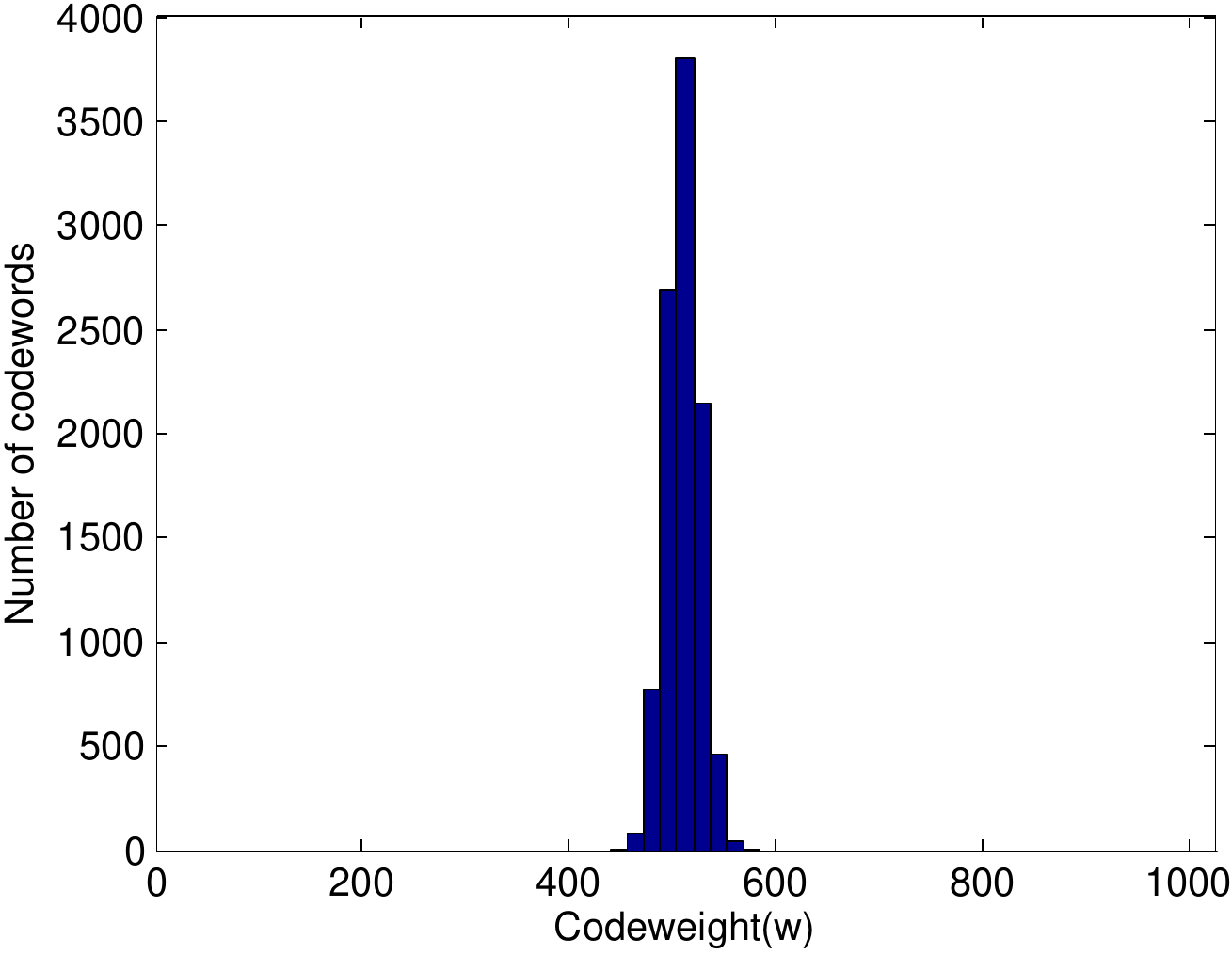}\label{fig_cw_25}}
\hfil
\subfloat[Code rate $R_c=1/2$]{\includegraphics[width=2.5in]{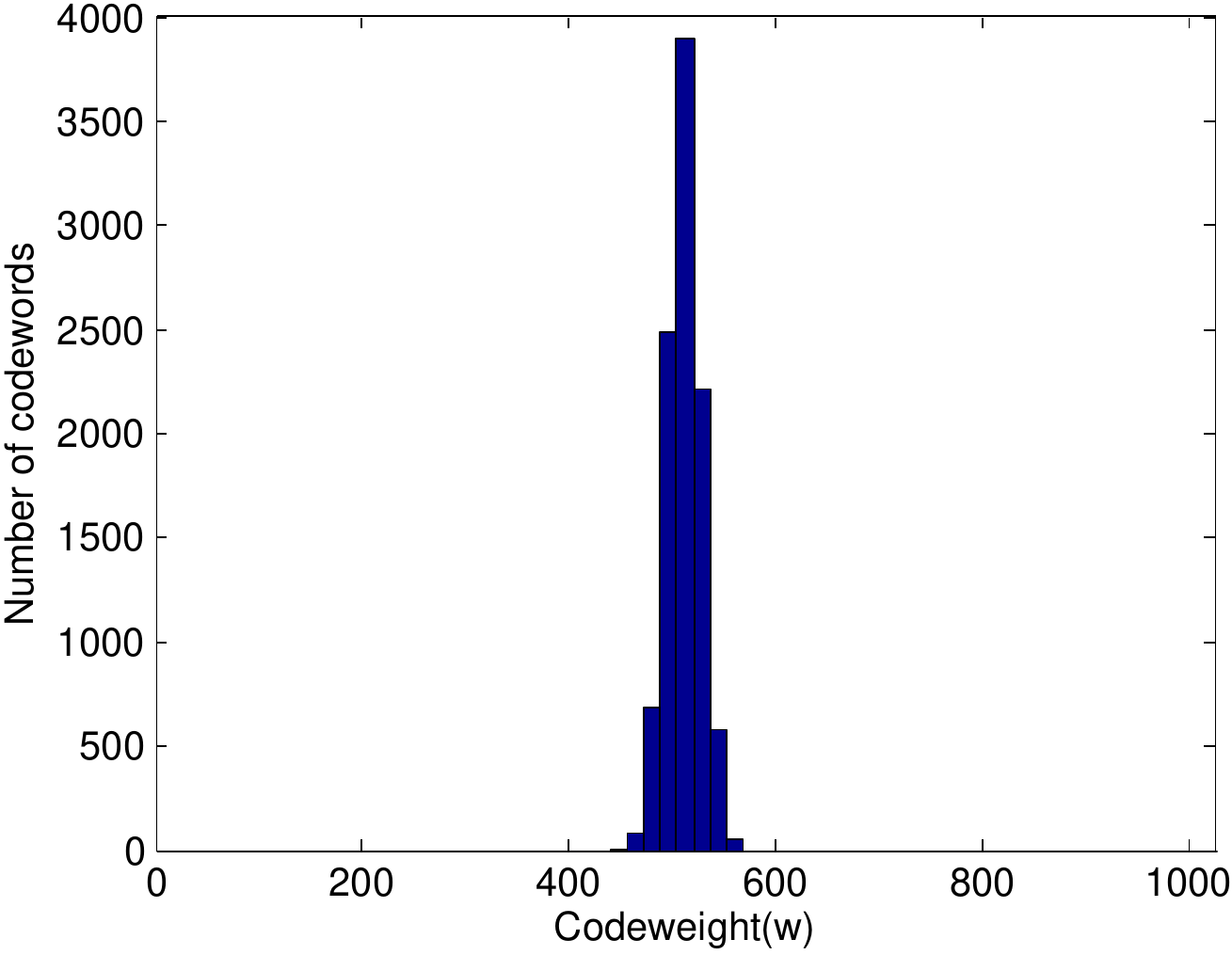}\label{fig_cw_50}}
\hfil
\subfloat[Code rate $R_c=3/4$]{\includegraphics[width=2.5in]{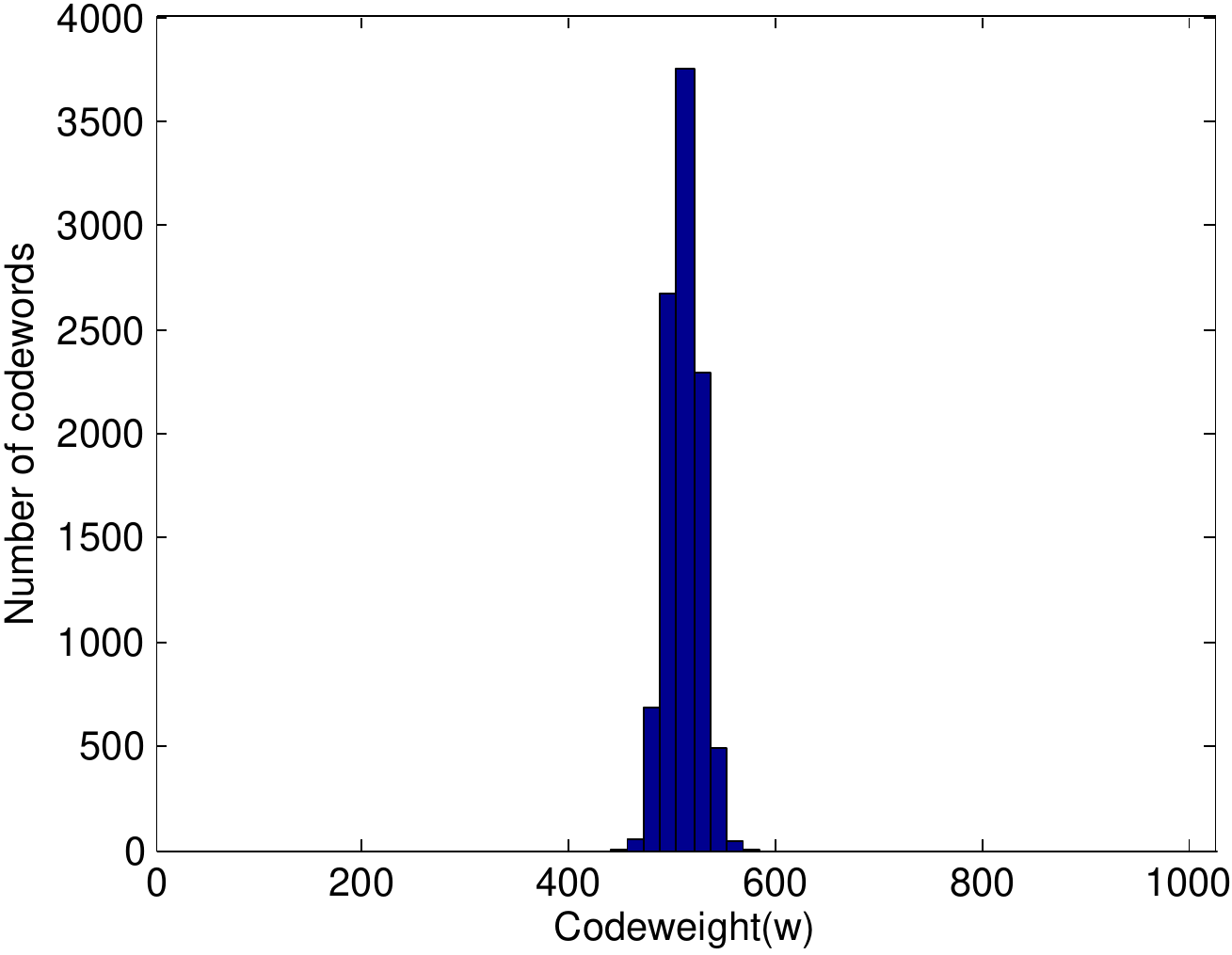}\label{fig_cw_75}}
\caption{Code weight distribution of polar codes with three code rates.}
\label{fig_cw}
\end{figure}

In the first experiment, we investigate the code weight distribution of polar codewords with code rate $1/4$, $1/2$ and $3/4$ for comparison.
For each code rate, 10,000 1024-bit codewords were generated by encoding 10,000 random input messages with polar encoder, and then the codewords' weights were statistically analyzed.
The results are shown as the histograms in Figure~\ref{fig_cw_25}, \ref{fig_cw_50} and~\ref{fig_cw_75}, correspondingly.

The first interesting finding in these histograms is that the weights of most codewords concentrate around 512, the half-length of the codewords, meaning that the numbers of 0's and 1's are almost equal, as well as the probabilities.
In details, there are about $86.4\%$ codewords have weights in the range of [488, 536], and about $13.6\%$ codewords have weights in the range of [448, 488] and [536, 560], while no codeword has a code weight less than 448 or larger than 560.
The mean code weight is $512$, and the standard deviation is $16$ (about $1.6\%$ of the codeword's length, $1024$).  
Considering the lowest optical clock rate of the IEEE VLC standard ($200 k$Hz) \cite{ieee2011std}, $1024$ symbols are transmitted during at most $5.12$ ms.
The standard deviation of the averaged dimming for 5.12 ms is $1.6\%$. This corresponds to a fluctuation of $50\pm 1.6\%$ at about 200 Hz, which is negligible compared to non-perceptible on-off pulses of $50\pm 50\%$~\cite{lee2012turbo}. 

This is one of the advantages that polar codes possess. 
Since the LEDs are turned on or off dependent on the data bits being 1 or 0 in OOK modulation, the weight of codeword represents the average light intensity over the duration of transmitting the codeword.
Therefore, the equal probability of 1's and 0's mean the balance of ON's and OFF's symbols, i.e., $50\%$ light intensity.
Based on this feature, LED light could be dimmed up or down for any desired brightness by simply inserting ON or OFF CSs without any bias constraint.

The second interesting finding is that all the three distributions are almost the same, regardless the different code rates.
This merit significantly enhances the flexibility of the proposed coding scheme to adapt to different channel qualities of VLC, while maintaining the stable intensity balance for dimming support.
For any BER of VLC channels, the proposed coding scheme could freely choose an appropriate code rate without the constraint of illumination.
The experiments in Section~\ref{sec_ecc} also illustrate this point.

More importantly, these two advantages could simplify the system design as the auxiliary coding techniques for weight balance such as scrambling,  DC-balance line coding and code puncturing are no longer needed in the proposed scheme for dimming support. 
Consequently, the proposed scheme could achieve a higher coding efficiency than the previous schemes, as analyzed in Section~\ref{sec_efficiency}.



\subsection{Short Run Length Property for Flicker Mitigation}
\label{sec_rll}
While the codewords' weight represent the average light intensity during transmitting data frame, the run length of 0's or 1's represents the intensity over a shorter period, which may exceed the persistence of the human eye and causes noticeable brightness change, i.e., flicker.
Therefore, the run length of codewords should be carefully limited to mitigate any potential flicker resulting from modulating the light sources for communication because flicker can cause negative/harmful physiological changes in humans~\cite{rajagopal2012ieee}.

\begin{figure}[!t]
\centering
\includegraphics[width=2.5in]{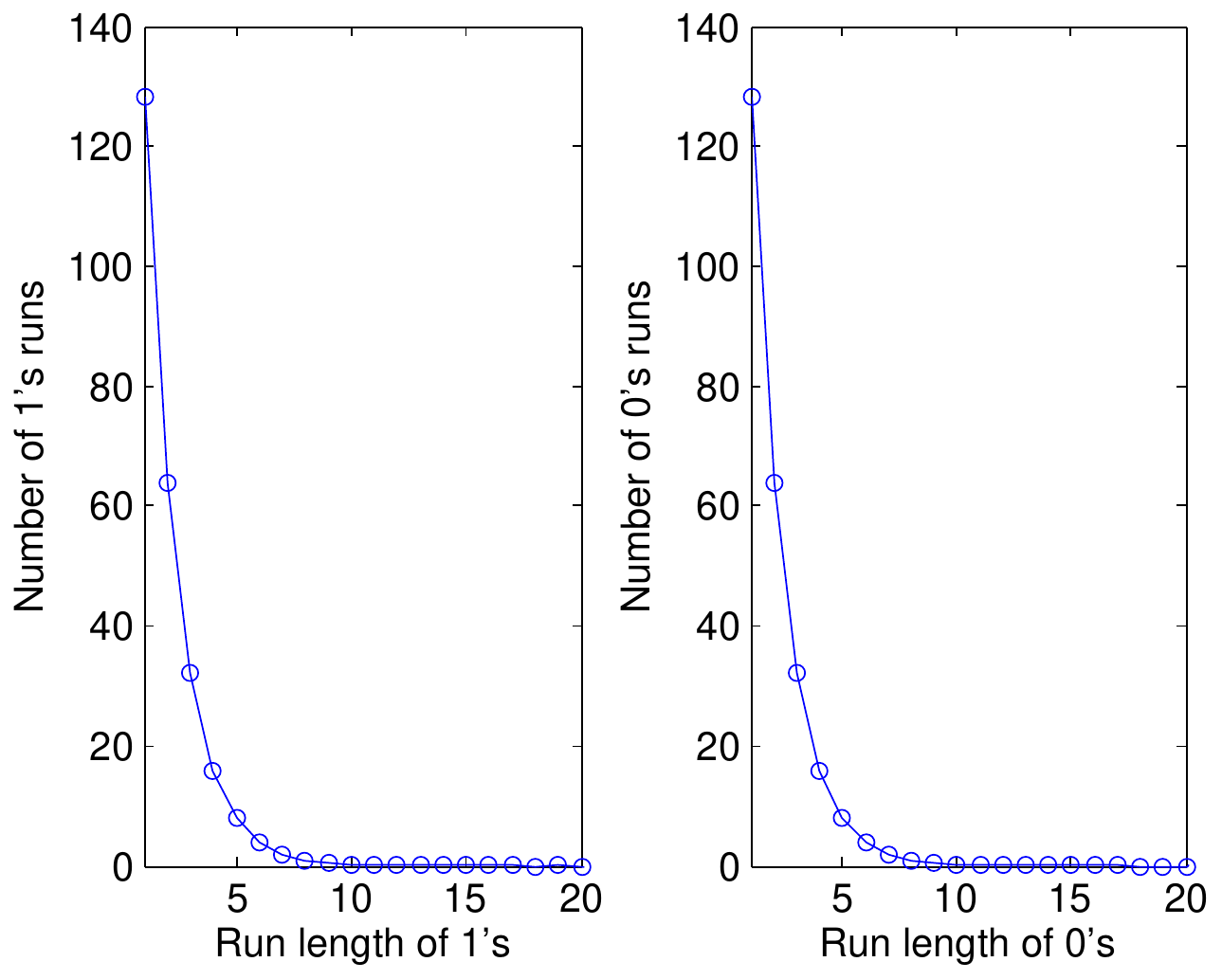}
\caption{Run Length Distribution of polar codes.}
\label{fig_rll}
\end{figure}

In this experiment, we further investigate the run length distribution of 10,000 1024-bit polar codewords and the results of statistics are shown in Figure~\ref{fig_rll}. 
With the increment of run length $l$, the average number ($n(l)$) of runs being of length $l$ drops with a fraction of $1/2$, e.g., $128, 64, 32 , \cdots$
Averagely, about 912 bits in each 1024-bit codewords are included in the runs shorter than 5 ($l<5$), which means that about $90\%$ bits in polar codewords hold the property of RLL.
The maximum run length in the 10000 tests is 20 and this case only occurs twice in the total $10000\times 1024$ bits. 
Even so, transmitting the 20 bits in the VLC system with the lowest optical clock rate ($200 k$ Hz) only requires 0.1 ms ($=20*1/200,000$), which is far less than the MFTP.

In short, polar codewords have an excellent advantage of short run length property for flicker mitigation.
And this advantage can further simplify the proposed VLC system design since RLL line coding is no longer needed in this system.
Moreover, the spectral efficiency of the proposed VLC system could also be increased by getting rid of the low-efficiency RLL line coding. 
 
\subsection{Overall Coding Efficiency}
\label{sec_efficiency}
Based on the above two advantages, the proposed coding scheme could achieve a higher overall coding efficiency than the other existing schemes.
The overall coding efficiency of the proposed scheme is compared in Table~\ref{tab_efficiency} with that in Ref.~\cite{kim2011novel} and Ref.~\cite{kim2015adaptive}, which uses RM, LDPC codes for dimmable VLC.
Noted that the lengths of the codewords in those schemes are 32 and 1200, respectively.

Recall that $\eta_{overall}=k/N_{frame}=k/(N+N_{cs})$. 
Therefore, when the dimming ratio is increased or decreased from $50\%$, $N_{cs}$ is increased and the overall coding efficiency is reduced, while the code rates are unchanged.
However, as the proposed scheme is inherently DC-balanced, it could have an arbitrary code rate and a slower decrement than LDPC codes-based scheme.
If code rate $R_c=0.5$ is used, the overall coding efficiency of the proposed scheme for $50\%$ dimming is $\eta_{overall}=0.5$, about twice higher than that of LDPC codes-based scheme.
Even for the other dimming ratios, the proposed scheme with $R_c=0.5$ code rate could have a coding efficiency about two-fold to that of LDPC codes-based scheme.
If code rate $R_c=0.75$ is used, the overall coding efficiency of the proposed scheme is about three times higher than that of LDPC codes-based scheme.
And compared with RM codes-based scheme, which also has a slow coding efficiency decrement versus dimming ratio but is limited to a specific code rate $log_2(N)/N$, the proposed scheme could support arbitrary code rates, which are more practical and higher than the fixed code rate of RM codes. 

\begin{table}[!t]
\renewcommand{\arraystretch}{1.1}
\caption{Comparison of coding efficiency}
\label{tab_efficiency}
\centering
\begin{tabular}{|c|c|c|c|}
\hline
Dimming & Scheme&Code rate& Coding efficiency\\
\hline
\multirow{3}{*}{$50\%$} & RM codes~\cite{kim2011novel}  & 0.156		& 0.156	\\
&LDPC codes~\cite{kim2015adaptive}   &  0.5	&	0.24	\\	 
& The proposed scheme & Any $R_c$	& $R_c$ \\	
\hline
\multirow{3}{*}{$25, 75\%$} & RM codes~\cite{kim2011novel}  & 0.25		& 0.125	\\
&LDPC codes~\cite{kim2015adaptive}   &  0.5	&	0.12	\\		 
& The proposed scheme & Any $R_c$	& $R_c/2$ \\	
\hline	
\multirow{3}{*}{$12.5, 87.5\%$} & RM codes~\cite{kim2011novel}  & 0.375		& 0.093	\\
&LDPC codes~\cite{kim2015adaptive} &  0.5	&	0.06	\\				 
& The proposed scheme & Any $R_c$	& $R_c/4$ \\	
\hline	
\end{tabular}
\end{table}

\subsection{Error Correction Performance}
\label{sec_ecc}
For VLC channel, we use the channel environment and parameters for determining noise variances in~\cite{komine2004fundamental,kim2011novel,kim2015adaptive}, in which the BER is also approximated by $OOK:BER=Q\sqrt{SNR_{rx}}$ for OOK modulation assuming a rectangular pulse shape whose duration equals the bit period. 
$SNR_{rx}$ denotes the signal-to-noise ratio of the received signal.

\begin{figure}[!t]
\centering
\subfloat[versus SNR]{\includegraphics[width=2.5in]{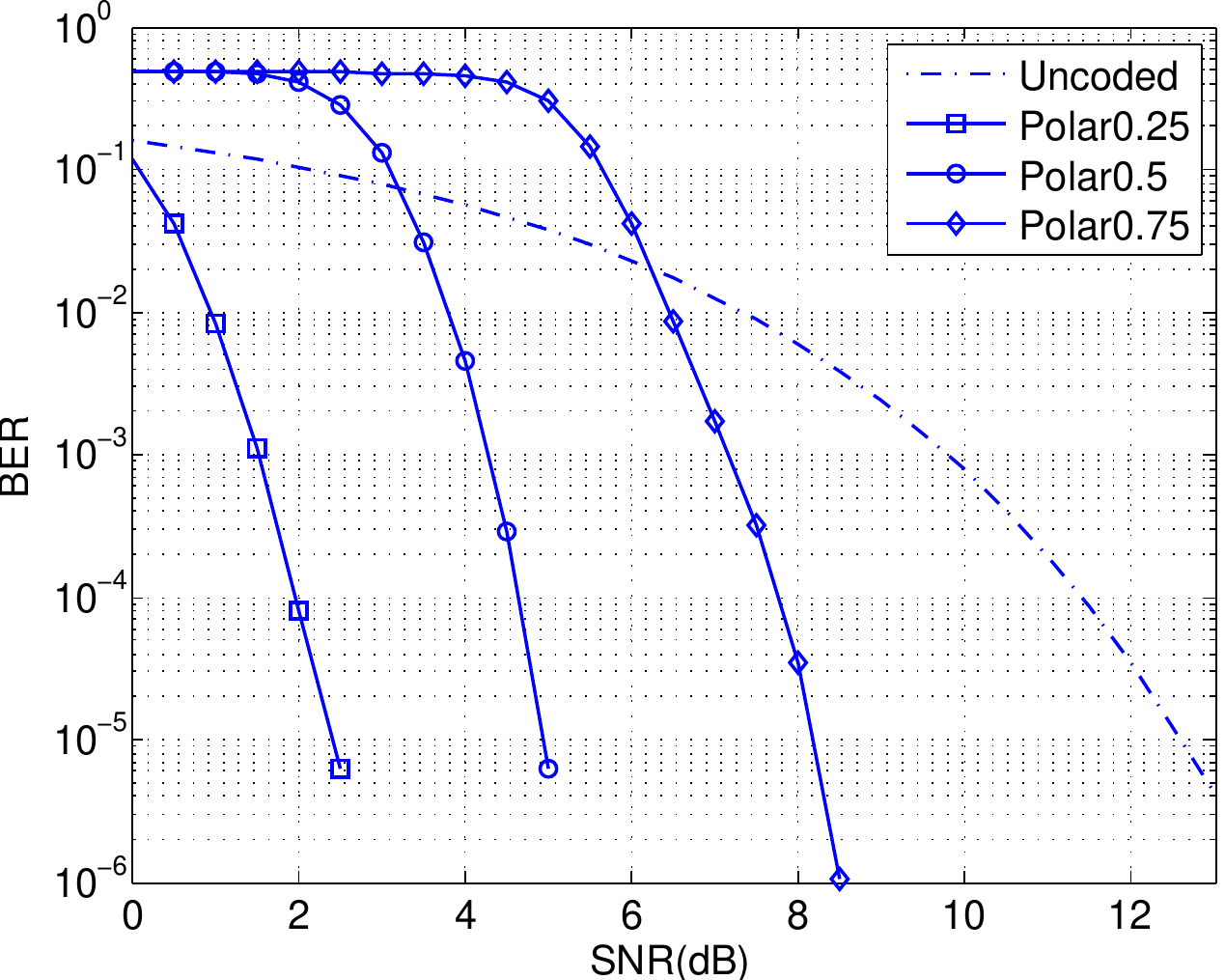}\label{fig_ecc_snr}}
\hfil
\subfloat[versus Eb/N0]{\includegraphics[width=2.5in]{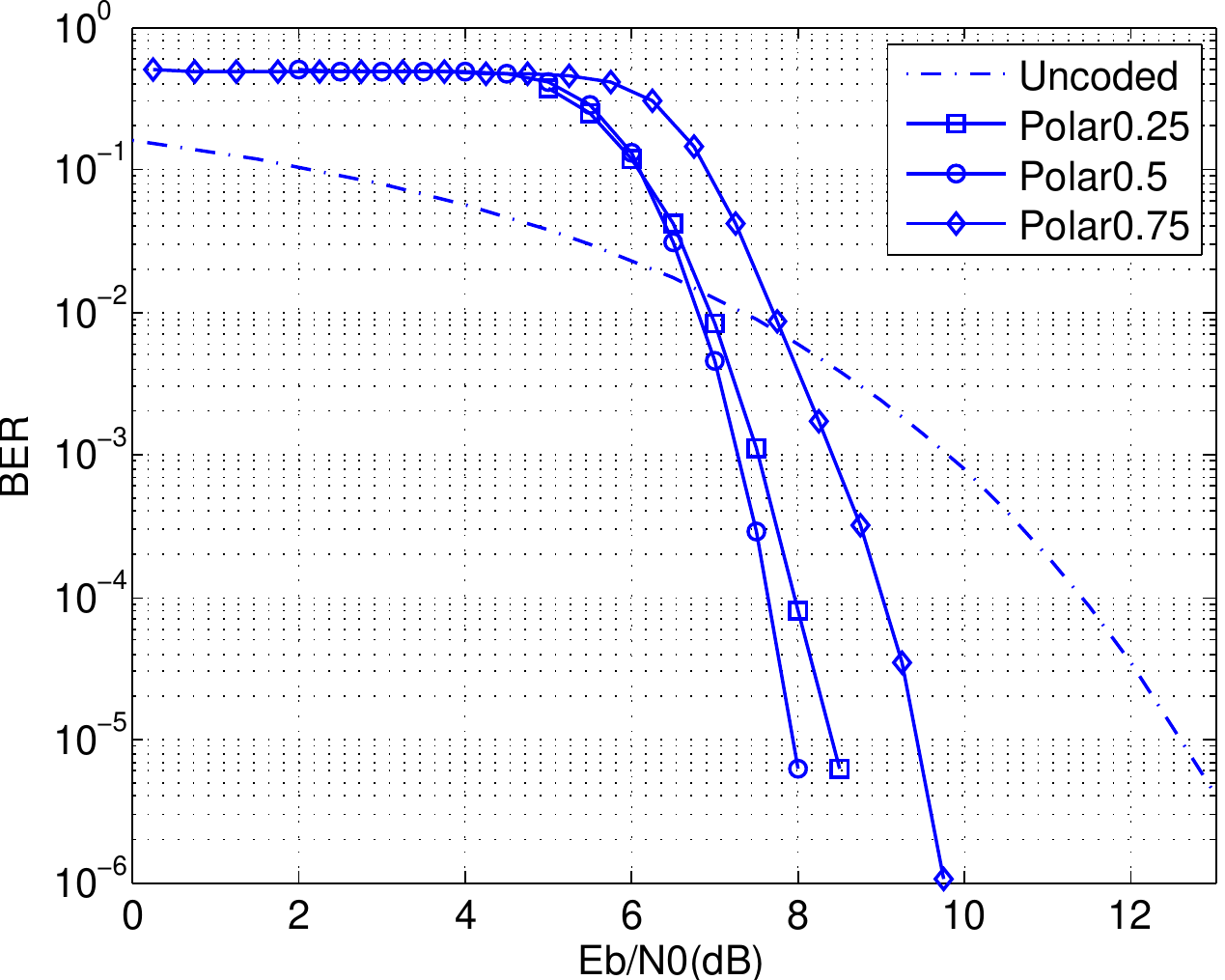}\label{fig_ecc_ebn0}}
\caption{Error correction performance of polar codes ($50\%$  dimming).}
\label{fig_ecc}
\end{figure}

Figure~\ref{fig_ecc_snr} shows the BER performance of the proposed scheme for $50\%$ dimming with three code rates varying with $SNR_{rx}$, compared with that of the uncoded system.
For each code rate and each SNR value, 10,000 1024-bit polar codewords were randomly generated and the transmission via AWGN channel was simulated for this experiment.
The numbers following ``Polar" in the legend of the figures denote the code rates of different curves.
As shown in the figure, using polar codes with code rate of 0.25, 0.5 and 0.75, the proposed VLC system can achieve a near error-free BER ($<10^{-5}$)~\cite{costello2007channel} when $SNR_{rx}$ is about 2.4 dB, 4.9dB and 8.2dB, respectively. 
As we discussed above, polar codes could guarantee the code weight balance regardless code rates, therefore, different code rates could be freely chosen to suit different application environments without impact on the dimming function.

To compare the BER performances of the proposed scheme and that of the other previous FEC schemes considering code rate and coding gain, the energy per bit to noise power spectral density ratio ($E_b/N_0$) is used for fair comparison and $E_b/N_0=SNR_{rx}/R_c$.

\begin{figure}[!t]
\centering
\includegraphics[width=2.5in]{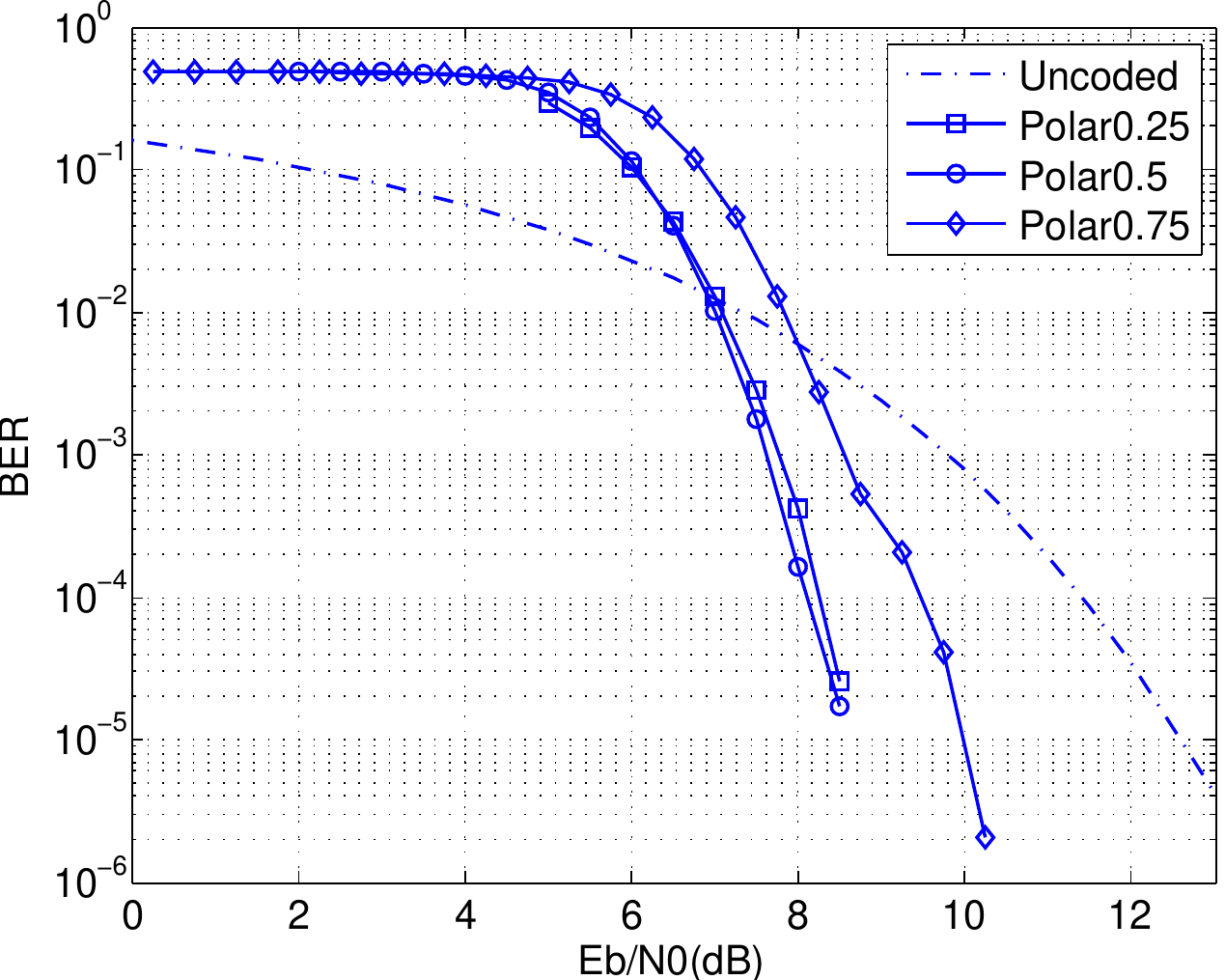}
\caption{Error correction performance of polar codes ($75\%$  dimming).}
\label{fig_ecc_ebn0_512}
\end{figure}

Figure~\ref{fig_ecc_ebn0} shows the BER performances of the proposed scheme for $50\%$ dimming with three code rates varying with $E_b/N_0$, compared with that of the uncoded system.
In terms of coding gain, the polar codes with three code rates have similar coding gain curve and the one with code rate $0.5$ reaches the near error-free BER at $E_b/N_0=7.9dB$, which is about 1 dB higher coding gain than that of modified RM codes~\cite{kim2011novel} and about 3 dB higher coding gain than that of RS(64,32) in IEEE standard 802.15.7~\cite{ieee2011std}.
For dimming ratio $25\%$ (or $75\%$), the proposed scheme can achieve about 1 dB higher than that of LDPC codes~\cite{kim2015adaptive},  as shown in Figure~\ref{fig_ecc_ebn0_512}.


\section{Conclusion}
\label{sec_conclusion}
VLC is the convergence of illumination and communication. 
Traditional FEC coding schemes need to be modified for VLC to provide two lighting related functionalities: dimming support and flicker mitigation. 
These schemes suffer from low coding efficiency and complicated coding structure due to the auxiliary coding procedures to provide these supports.
In this paper, a capacity-achieving and flicker-free FEC coding scheme using polar codes is proposed for dimmable VLC.
Experimental results show that the proposed scheme has outstanding advantages in code weight balance, short run length property, higher coding efficiency and better error correction performance.
Therefore, the dimmable VLC system with the proposed FEC coding scheme can achieve not only a simplified system design but also a high coding efficiency with capacity-achieving error correction performance.


%



\section*{Acknowledgment}
This work was partially supported by China State Scholarship Fund(No. 201506785014), National Natural Science Foundation of China (No. 61401176, 61402136, 61672014, 61361166006), Natural Science Foundation of Guangdong Province (No. 2014A030310205, 2014A030313697, 2016A030313090), Science and technology projects of Guangdong Province (2014B010120002, 2016A010101017), Project of Guangdong High Education  (YQ2015018), and NSFC/RGC Joint Research Scheme (N\_HKU 72913), Hong Kong.

\ifCLASSOPTIONcaptionsoff
  \newpage
\fi



%

\bibliographystyle{IEEEtran}
\bibliography{vlcpolar_jphoto}

%
%

%

\begin{IEEEbiographynophoto}{Junbin Fang} received the Ph.D. degree from South China Normal University, China, in 2008. He is currently an Associate Professor with the Department of Optoelectronic Engineering, Jinan University and a visiting professor with The Edward S. Rogers Sr. Department of Electrical \& Computer Engineering Department, University of Toronto. His research focuses on visible light communication, channel coding, communication and information security.
\end{IEEEbiographynophoto}
 \vspace{-10 mm}

\begin{IEEEbiographynophoto}{Zhen Che} received his BEng degree in photoelectric information engineering from Jinan University, Guangzhou, Guangdong, China, in 2011. He is now a PhD student at the College of Information Science and Technology, Jinan University, Guangzhou, Guangdong, China. His research interests include visible light communication, LED devices and optical simulation.
\end{IEEEbiographynophoto}
 \vspace{-10 mm}

\begin{IEEEbiographynophoto}{Xiaolong Yu} received his BEng degree in electronics science and technology from East China Institute of Technology, Jiangxi, China, in 2014. He is now a master student at the Department of Optoelectronic Engineering, Jinan University, Guangzhou, Guangdong, China. His research interests include visible light communication and channel coding.
\end{IEEEbiographynophoto}
 \vspace{-10 mm}
 
\begin{IEEEbiographynophoto}{Zhe Chen} is a professor in the Department of Optoelectronic Engineering, Jinan University, China. He got the PhD in physical electronics from Tsinghua University, master’s degree in military optics and bachelor’s degree in applied physics from National University of Defense Technology. His research interests include visible light communication, micro/nano optical devices, photoelectric information, optical fiber devices, optical fiber communication and sensing technology, optical system design and application.
\end{IEEEbiographynophoto}
 \vspace{-10 mm}

\begin{IEEEbiographynophoto}{Zoe L. Jiang} received the Ph.D. degree from The University of Hong Kong, Hong Kong, in 2010. She is currently an Assistant Researcher with School of Computer Science and Technology, Harbin Institute of Technology Shenzhen Graduate School, China. Her research interests include wireless communication security, cryptography and cloud security.
\end{IEEEbiographynophoto}
 \vspace{-10 mm}

\begin{IEEEbiographynophoto}{Siu-Ming Yiu} is currently an Associate Professor in the Department of Computer Science of the University of Hong Kong. He is also the leader of the applied cryptographic research group and an Associate Programme Director of the Information Security stream of the MSc(CS) programme. His resaerch interests include cryptography, computer security, and bioinformatics.
\end{IEEEbiographynophoto}
 \vspace{-10 mm}

\begin{IEEEbiographynophoto}{Kui Ren}is a professor of Computer Science and Engineering and the director of UbiSeC Lab at State University of New York at Buffalo (UB). He received his PhD degree from Worcester Polytechnic Institute. Kui's current research interest spans Cloud \& Outsourcing Security, Wireless \& Wearable Systems Security, and Mobile Sensing \& Crowdsourcing. His research has been supported by NSF, DoE, AFRL, MSR, and Amazon. He received UB Exceptional Scholar Award for Sustained Achievement in 2016, UB SEAS Senior Researcher of the Year Award in 2015, Sigma Xi/IIT Research Excellence Award in 2012, and NSF CAREER Award in 2011. Kui has published 170 peer-review journal and conference papers and received several Best Paper Awards including IEEE ICNP 2011. He currently serves as an associate editor for IEEE Transactions on Dependable and Secure Computing, IEEE Transactions on Mobile Computing, IEEE Wireless Communications, IEEE Internet of Things Journal, and IEEE Transactions on Smart Grid. Kui is a Fellow of IEEE, a Distinguished Lecturer of IEEE, a member of ACM, and a past board member of Internet Privacy Task Force, State of Illinois.
\end{IEEEbiographynophoto}
 \vspace{-10 mm}

\begin{IEEEbiographynophoto}{Xiaoqing Tan} received her Ph.D in Applied Mathematics at Sun Yet-sen University. She is an Associate Professor at the Department of Mathematics, Jinan University. Her research interests include issues related to quantum cryptography, classical cryptography and error-correcting code theory. She is an author of a great deal of research studies published at international and national journals, and conference proceedings.
\end{IEEEbiographynophoto}








\end{document}